\documentclass[aps,prb,amsmath,twocolumn,groupedaddress,superscriptaddress,floatfix,citeautoscript,showpacs]{revtex4-1}

\usepackage[dvips]{graphicx}

\usepackage{subfigure}

\begin{document}

\title{Anomalous ideal tensile strength of ferromagnetic Fe and Fe-rich alloys}

\author{Xiaoqing Li}
\email{xiaoqli@kth.se}
\affiliation{Applied Materials Physics, Department of Materials Science and Engineering, Royal Institute of Technology, Stockholm SE-10044, Sweden}
\author{Stephan Sch\"onecker}
\email{stesch@kth.se}
\affiliation{Applied Materials Physics, Department of Materials Science and Engineering, Royal Institute of Technology, Stockholm SE-10044, Sweden}
\author{Jijun Zhao}
\email{zhaojj@dlut.edu.cn}
\affiliation{Key Laboratory of Materials Modification by Laser, Ion and Electron Beams (Dalian University of Technology), Ministry of Education, Dalian 116024, China}
\author{B\"{o}rje Johansson}
\affiliation{Applied Materials Physics, Department of Materials Science and Engineering, Royal Institute of Technology, Stockholm SE-10044, Sweden}
\affiliation{Department of Physics and Astronomy, Division of Materials Theory, Uppsala University, Box 516, SE-75120, Uppsala, Sweden}
\author{Levente Vitos}
\affiliation{Applied Materials Physics, Department of Materials Science and Engineering, Royal Institute of Technology, Stockholm SE-10044, Sweden}
\affiliation{Department of Physics and Astronomy, Division of Materials Theory, Uppsala University, Box 516, SE-75120, Uppsala, Sweden}
\affiliation{Research Institute for Solid State Physics and Optics, Wigner Research Center for Physics, Budapest H-1525, P.O. Box 49, Hungary}

\date{\today}

\begin{abstract}
Within the same failure mode, iron has the lowest ideal tensile strength among the transition metals crystallizing in the body-centered cubic structure. Here, we demonstrate that this anomalously low strength of Fe originates partly from magnetism and is reflected in unexpected alloying effects in dilute Fe(\emph{M}) (\emph{M} = Al, V, Cr, Mn, Co, Ni) binaries. We employ the structural energy difference and the magnetic pressure to disentangle the magnetic effect on the ideal tensile strength from the chemical effect. We find that the investigated solutes strongly alter the magnetic response of the Fe host from the weak towards a stronger ferromagnetic behavior, which is explained based on single-particle band energies.
\end{abstract}

\maketitle

Improving the strength and ductility of materials is an eternal challenge in materials design.
The strength of most structural materials is determined by the complex micro-structural properties associated with defects, such as vacancies, dislocation networks, and grain boundaries. A realistic description of strength involves accurate modeling of the dislocation activity for long periods of time, which is an enormous task for \emph{ab initio} methods.
However, the mechanical strength of solids is bounded from above and the limit is referred to as the ideal strength~\cite{Kelly:1986}.
The ideal tensile strength (ITS) is the stress at which a perfect crystal in tension becomes unstable with respect to an infinitesimal homogeneous strain. The ITS has been accepted as an essential intrinsic mechanical parameter of single crystal materials~\cite{inherent:property,study:8,Li:2007}.
The ideal strength connects aspects of chemical bonding and crystal symmetry with the mechanical properties of ideal lattices, such as the failure mode~\cite{Qi:2014} or common slip systems~\cite{inherent:property},
and is involved in fracture theory and the nucleation of defects~\cite{Thomson:1986,*Jokl:1980,Kelly:1986}. The ideal strength can be approached in systems with very low defect density like whiskers or thin films, and in graphene~\cite{Kelly:1986,Krenn:2002,Lee:2008}.

Considerable progress has been made to understand the behavior of elemental solids~\cite{study:9,Clatterbuck:2003,Cerny:2013}, compounds~\cite{Chen:2007,Zhang:2004,Blase:2004,*Blase:2,jiang:2013}, and ordered alloys~\cite{study:1} at the limit of strength. Although steels are well-studied structural materials, surprisingly little is known about their ideal strengths~\cite{Guo:2001,*Morris:2001b}. Previous \emph{ab initio} studies focused on elemental Fe as the basic ingredient to steel~\cite{inherent:property,Cerny:2007,Sob:2004,Cerny:2013,Cerny:2010}.
Accordingly, ferromagnetic Fe fails in tension by cleavage of the $\{001\}$ atomic planes at an attainable strength $\sim 55\,\%$ lower than those of the other body-centered cubic (bcc) metals with the same failure mode (Mo and W)~\cite{Cerny:2007}. The bcc refractory elements V, Nb, and Ta possess comparable low ITSs to Fe, but they fail under tensile stress due to a shear instability rather than by cleavage~\cite{study:8}.
Restricting the failure in tension to cleavage of the $\{001\}$ planes, the ITS of Fe turns out to be the lowest among \emph{all} bcc transition metals~\cite{Mo:1,study:8,inherent:property,Cerny:2010}.

Against this background, several fundamental issues call for in-depth investigation: Why is the ITS of Fe so much lower than those of Mo and W?  Can the ideal strength of Fe be increased by alloying? Will the failure mode in Fe-alloys be changed? What role does magnetism play in the attainable ITS of Fe and its alloys?

In this paper, we employ \emph{ab initio} alloy theory to shed light onto the above questions. Studying pure Fe and six Fe-based binary alloys, we predict that the ITS of Fe can be significantly altered by alloying. We demonstrate that the ITS of Fe is anomalously low due to the weak ferromagnetic behavior.
We give evidence that not only the late $3d$ metals but also a small amount of early $3d$ metal or Al enhance the stability of the ferromagnetic order in Fe, and this change in the magnetic behavior dominates the alloying effects on the ITS of Fe-rich alloys.

The ITS of Fe ($\sigma_\text{m}$) is the first maximum of the stress-strain curve, $\sigma(\epsilon)=\frac{1+\epsilon}{\Omega(\epsilon)}\frac{\partial E(\epsilon)}{\partial \epsilon}$, with corresponding maximum strain ($\epsilon_{\text{m}}$) upon uniaxial loading along the $\langle 001\rangle$ direction. Here, $\Omega(\epsilon)$ is the relaxed volume at strain $\epsilon$. The tensile stress was determined by incrementally straining the crystal and taking the derivative of the computed total energy $E(\epsilon)$ with respect to $\epsilon$. At each value of the strain, the two unit cell lattice vectors perpendicular to the $\langle 001\rangle$ direction were relaxed allowing for a possible symmetry lowering deformation relative to the initial body-centered tetragonal symmetry (bct, lattice parameters $a$ and $c$). Previous studies of the magnetic order of bct Fe~\cite{idealandmagnetic:2002,Tsetseris:2005,Friak:2001} showed that Fe remains ferromagnetic in the part of the ($a$, $c$)-configuration space
corresponding to $\langle
001\rangle$ uniaxial tension with $\epsilon\leq\epsilon_{\text{m}}$. Accordingly, all calculations were performed for ferromagnetic Fe matrix.
\begin{table}[thb]
\centering
\caption{\label{table:ITS}The present ideal tensile strength ($\sigma_{\text{m}}$) and the corresponding strain ($\epsilon_{\text{m}}$) under $\langle 001 \rangle$ loading for ferromagnetic bcc Fe compared with the available literature data for Fe, Mo and W.}

\begin{ruledtabular}
\begin{tabular}{ccc}
element   & $\sigma_{\text{m}}$ (GPa) & $\epsilon_{\text{m}} (\%)$ \\
 \hline
    Fe & 12.6 & 14.1\\
     &12.6~\cite{inherent:property}, 12.4~\cite{Cerny:2007} & 15~\cite{inherent:property}, 16~\cite{Cerny:2007} \\
       & 12.4~\cite{Liu}, 12.7~\cite{Sob:2004} & 14~\cite{Liu}, 15~\cite{Sob:2004} \\
     Mo & 28.3~\cite{Cerny:2007} & 12~\cite{Cerny:2007} \\
     W  & 28.9~\cite{Cerny:2007} & 13~\cite{Cerny:2007} \\
     \end{tabular}
\end{ruledtabular}
\end{table}

The adopted first-principles method is based on density-functional theory
as implemented in the exact muffin-tin orbitals method~\cite{EMTO:1,*EMTO:2,*EMTO:3} with exchange-correlation parameterized by Ref.~\onlinecite{PBE,*Perdew:1996E}. The problem of disorder was treated within the coherent-potential approximation~\cite{cpa:1,*cpa:3}, and the total energy was computed via the full charge-density technique~\cite{EMTO:1,*EMTO:2,*EMTO:3}.

Our calculated ITS for Fe is in close agreement with previous assessments (Table~\ref{table:ITS}). We found that a bifurcation from tetragonal to orthorhombic symmetry occurs at  $\epsilon_{\text{orth}}=17\,\%$ (due to a vanishing shear modulus), i.e., well above $\epsilon_{\text{m}}=14.1\%$. This result is in accordance with Ref.~\onlinecite{inherent:property}, where the branching was reported to occur at $18$\% strain. Compared to the ITSs of Mo and W (Table~\ref{table:ITS}), which possess the same failure mode as Fe, the ITS of Fe is anomalously low.
\begin{figure}[htb]
\begin{center}
\resizebox{0.95\columnwidth}{!}{\includegraphics[clip]{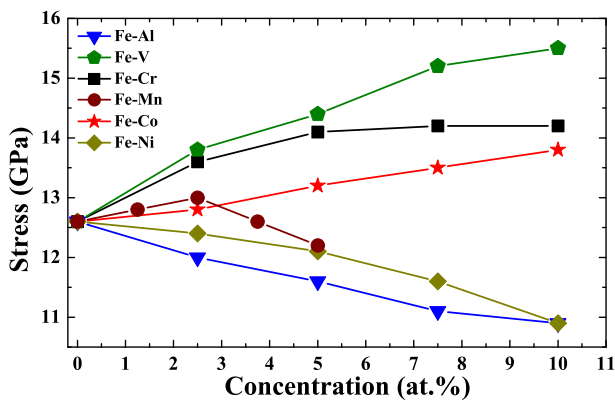}}%
\caption{\label{fig:ITS}(Color online) The ITS ($\sigma_\text{m}$) of ferromagnetic bcc Fe$_{1-x}M_{x}$ alloys as a function of concentration.}
\end{center}
\end{figure}

Turning to Fe$_{1-x}M_{x}$ (\emph{M} = Al, V, Cr, Mn, Co, or Ni) solid solutions, we studied the alloying effect on the ITS in the concentration interval $0\le x\le0.1$ (for Mn $0\le x\le0.05$). The selected solute atoms are common in commercial steel alloys. The calculated ITS is found to increase with V, Cr and Co and decrease with Al and Ni addition to Fe (Fig.~\ref{fig:ITS}).
Manganese shows a small but non-monotonous alloying behavior. When accounting for the possibility of an orthorhombic branching away from the tetragonal deformation path, we found that for the present binaries and concentrations, the branching occurs at strains larger than $\epsilon_{\text{m}}(x)$ corresponding to the ITS of Fe$_{1-x}M_x$. Hence all Fe-alloys considered here are predicted to fail by cleavage under $\langle 001 \rangle$ loading.

In the following, we analyze the alloying effect on the ITS of Fe$_{1-x}M_{x}$ starting from a model based on structural energy differences (SEDs)~\cite{Xiaoqing:2013}. Although it turns out that the SED model fails for $\sigma_{\text{m}}$, we find that this model accounts for the alloying effect on the auxiliary ITSs obtained either for constant-volume deformation ($\sigma^{\Omega}_{\text{m}}$) or for fixed-magnetic moment along the relaxed loading path ($\sigma^{\mu}_\text{m}$). We show that the difference between the auxiliary and the full ITSs, viz. $\Delta\Sigma^{\Omega/\mu}\equiv \sigma^{\Omega/\mu}_{\text{m}}-\sigma_{\text{m}}$, correlates well with the excess magnetic pressure ($\Delta P_{\text{mag}}$) that develops upon lattice distortion. Finally, the difference in the magnetic properties of Fe and Fe$_{1-x}M_{x}$ are elucidated on the basis of their electronic structure.

\begin{figure}[tbh]
\begin{center}
\resizebox{0.95\columnwidth}{!}{\includegraphics[clip]{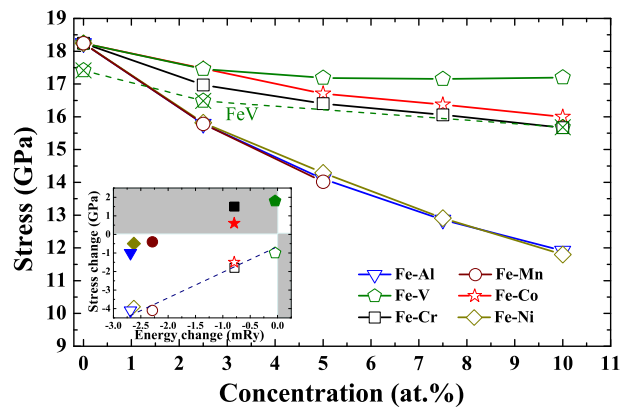}}
\caption{\label{fig:SED}(Color online) The constrained ITSs of ferromagnetic bcc Fe$_{1-x}M_{x}$ alloys at a constant-volume deformation ($\sigma^{\Omega}_{\text{m}}$) and with fixed-magnetic moment along the relaxed loading path ($\sigma^{\mu}_{\text{m}}$, crossed-out pentagons).
The inset shows the change in the ITS versus the change in the fcc-bcc SED for a concentration increase from $x_\text{i}=0.00$ to $x_\text{f}=0.05$. Filled symbols and open symbols denote the data for the fully relaxed loading paths ($\sigma_\text{m}$, Fig.~\ref{fig:ITS}) and the constrained-volume deformations paths ($\sigma^{\Omega}_\text{m}$), respectively. The SED model assumes that the data points are located along a straight line (dashed line) in the unshaded area.}
\end{center}
\end{figure}

The SEDs have often been used to explain alloying trends of various physical parameters~\cite{Xiaoqing:2013,Wills:1992,*Souvatzis:2004}. According to the SED model for the ITS~\cite{Xiaoqing:2013}, the change in $\sigma_{\text{m}}(x)$ with concentration variation from $x_\text{i}$ to $x_\text{f}$ should be proportional to the corresponding change in SED, viz. $[\sigma_\text{m}(x_\text{f})-\sigma_{\text{m}}(x_\text{i})] \propto [\Delta E_{\text{SED}}(x_\text{f})-\Delta E_{\text{SED}}(x_\text{i})]$, where $\Delta E_{\text{SED}}(x)\equiv E_{\text{fcc}}(x)-E_{\text{bcc}}(x)$ is the SED between the face-centered cubic (fcc) and bcc lattices both of them taken at the bcc equilibrium volume. Surprisingly, no such correlation could be established for the present Fe-alloys. This is illustrated in the inset of Fig.~\ref{fig:SED} (filled symbols), where the SEDs were computed assuming ferromagnetic order for both lattices.
On the other hand, the SED model provides the possibility to isolate that part of $\sigma_\text{m}(x)$ which correlates with the trend of SED. To this end, we consider $\sigma^{\Omega}_\text{m}$ obtained by fixing the volumes along the deformation paths to the respective bcc equilibrium volumes. Monitoring $\sigma^{\Omega}_\text{m}$ (Fig.~\ref{fig:SED}), we find that the change in $\sigma^{\Omega}_\text{m}(x)$ follows the change in $\Delta E_{\text{SED}}(x)$. This is demonstrated in the inset of Fig.~\ref{fig:SED} (open symbols).

Comparing $\sigma^{\Omega}_\text{m}$ (Fig.~\ref{fig:SED}) with $\sigma_\text{m}$ (Fig.~\ref{fig:ITS}), we realize that the ITS of pure Fe is the most significantly increased by the constant-volume constraint. This is to a much lesser extent the case for Fe-alloys, for which $\Delta\Sigma^{\Omega}$ are considerably smaller than for pure Fe. To reveal the origin of these differences, below we identify another auxiliary ITS that follows a similar trend as $\sigma^{\Omega}_\text{m}(x)$. Investigating the magnetic structures of Fe along the relaxed and constant-volume loading paths, we observe that in the relevant strain interval $\epsilon$:(0 - $\epsilon_\text{m}$) the value of the magnetic moment ($\mu$) of Fe is more strongly affected by the volume increase accompanying the tensile deformation ($\mu$ increases by $0.30\,\mu_{\text{B}}$ along the fully relaxed loading path) than by the lattice distortion ($\mu$ increases by $0.10\,\mu_{\text{B}}$ along the constant-volume deformation path).
Furthermore, we find that both of these changes of $\mu$ are substantially diminished by alloying. We give evidence how magnetism affects the ITS by considering $\sigma^{\mu}_\text{m}$ obtained by constraining the magnetic moment along the previously determined fully-relaxed strain paths to their respective ground state (bcc) values. According to $\sigma^{\mu}_\text{m}$ (exemplary shown in Fig.~\ref{fig:SED} for Fe$_{1-x}V_{x}$, crossed-out symbols), we observe that fixing the magnetic moment but taking into account structural relaxations yields very similar ITS for Fe and produces essentially the same alloying effect on the ITS as fixing the volume but allowing for the relaxation of the magnetic moments. Since for the latter deformation, the change of $\mu$ is small (as discussed above), both $\sigma^{\Omega}_\text{m}$ and $\sigma^{\mu}_\text{m}$ reflect the hypothetical case if Fe were a strong ferromagnet, for which the magnetic moment would not be sensitive to the atomic environment. Hence, we attribute the large values of $\Delta\Sigma^{\Omega/\mu}$ for pure Fe to the weak ferromagnetism, which is however very sensitively modified by alloying.

\begin{figure}[thb]
\begin{center}
\resizebox{0.95\columnwidth}{!}{\includegraphics[clip]{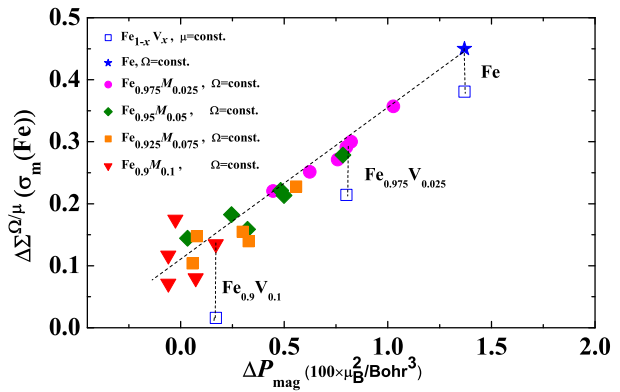}}%
\caption{\label{fig:correlation}(Color online) Correlation between the stress change ($\Delta\Sigma^{\Omega/\mu}$) in units of the ITS of Fe and the magnetic pressure change ($\Delta P_{\text{mag}}$) for the constant-volume deformation (filled symbols) and for the fixed-magnetic moment deformation along the relaxed loading path (open symbols). The dashed line indicates the linear correlation between $\Delta\Sigma^{\Omega}$ and $\Delta P_{\text{mag}}$.}
\end{center}
\end{figure}

To quantify the impact of magnetism on the ITS of Fe and Fe-alloys we make use of the concept of magnetic pressure in itinerant magnets.
We argue that if a positive excess magnetic pressure develops in the crystal upon constrained distortion, this excess pressure leads to a reduction of the total stress when the magnetic moment (or the volume) is released. Within the Stoner model, the magnetic pressure ($p_{\text{mag}}$) is estimated by $p_{\text{mag}}\propto k\mu^2/\Omega$, where $k$ stands for a positive proportionality factor that depends on potential parameters~\cite{Andersen:1977,Punkkinen:2011}. Accordingly, here we introduce the excess magnetic pressure as $\Delta P_{\text{mag}} \sim \mu^2_{\text{m}}/\Omega_{\text{m}}-\mu^2_{0}/\Omega_{0}$, evaluated at the ITS (subscript '$\textrm{m}$') with respect to the magnetic pressure present already in the ground state (subscript '$0$'). $\Delta P_{\text{mag}}$ describes the additional magnetic pressure in the lattice corresponding to the increase of the magnetic moment upon lattice distortion.
The two ITS enhancements $\Delta\Sigma^{\Omega}$ and $\Delta\Sigma^{\mu}$ against $\Delta P_{\text{mag}}$ are plotted in Fig.~\ref{fig:correlation} in units of the ITS of pure Fe.
The correlation between $\Delta\Sigma^{\Omega/\mu}$ and $\Delta P_{\text{mag}}$ is found to be very good for $x\le 0.075$ (with some scatter for $x=0.1$), meaning that larger excess magnetic pressure indeed yields larger difference between $\sigma^{\Omega/\mu}_{\text{m}}$ (Fig.~\ref{fig:SED}) and $\sigma_\text{m}$ (Fig.~\ref{fig:ITS}).
Due to its weak ferromagnetism, Fe exhibits the largest excess magnetic pressure, which induces $45\%$ ($38\%$) stress increase when constraining volume (magnetic moment).
In Fe$_{1-x}M_{x}$, the most important effect of alloying is that the excess magnetic pressure and thus $\Delta\Sigma^{\Omega/\mu}$ are gradually reduced with increasing $x$.
Alloys with 10\,\% solute concentration possess approximately zero excess pressure, meaning that there is only a small magnetic contribution to their ITSs.

The analysis above based on the excess magnetic pressure explains the differences between the ITSs shown in Fig.~\ref{fig:ITS} and Fig.~\ref{fig:SED}. Now one should ask why the magnetic moments of Fe-alloys (and the magnetic pressures) are much less sensitive to the atomic environment compared to that of pure Fe. We seek for a plausible explanation by considering how the single-particle band energies ($e$) are affected when an additional magnetic moment is induced in the lattice.
Employing the rigid band model and the force theorem~\cite{Springford:1980,*Skriver:1985}, we express the energy change by the change in the band energy ($\Delta E_{\text{band}}$) when the magnetic moment is increased by $\Delta \mu=\mu-\mu_0$ relative to the equilibrium moment ($\mu_0$).
Accordingly, $\Delta E_{\text{band}}\equiv E_{\text{band}\uparrow} + E_{\text{band}\downarrow}$ $= \int_{e_{\text{F}}}^{e_{\uparrow}(\mu)}(e'-e_{\text{F}})N_{\uparrow}(e')\text{d}e'+ \int^{e_{\downarrow}(\mu)}_{e_{\text{F}}}(e'-e_{\text{F}})N_{\downarrow}(e')\text{d}e'$, is the band energy change produced by transferring $(\mu-\mu_0)/(2\mu_{\text{B}})$ electrons from the minority-spin band ($N_{\downarrow}(e)$) below the Fermi level ($e_{\text{F}}$) to the majority-spin band ($N_{\uparrow}(e)$) above $e_{\text{F}}$. All required quantities were found from the calculated spin-polarized electronic density of states (DOS) at the corresponding bcc equilibrium volumes of alloys.

\begin{figure}[thb]
\begin{center}
\resizebox{0.95\columnwidth}{!}{\includegraphics[clip]{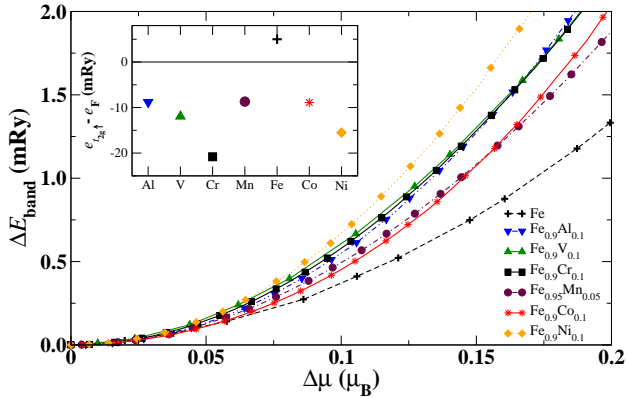}}%
\caption{\label{fig:DOS}(Color online) Band energy cost for an increase of the magnetic moment beyond its equilibrium value based on a rigid band analysis of the respective bcc total DOSs.
The inset shows the position of the $t_{2\text{g}\uparrow}$ shoulder in the majority-spin band relative to the Fermi energy as extracted from the equilibrium DOSs.
}
\end{center}
\end{figure}

Figure~\ref{fig:DOS} illustrates that the band energy cost for an increase of $\mu$ is lower for bcc Fe than for all present binaries with $x=0.1$ ($x=0.05$ in the case of Mn).
Therefore, an increase of the magnetic moment, e.g., as a result of a lattice distortion, is energetically much more favorable in pure Fe than in Fe alloys.
In other words, the magnetic moment (and thus the magnitude of the magnetic pressure) is much more sensitive to the atomic environment in pure Fe than in Fe-alloys.

Considering the spin-resolved changes of the band energy (not shown), we find that for pure Fe, the contribution of $N_\downarrow$ to $\Delta E_{\text{band}}$ is approximately twice as large as the contribution of $N_\uparrow$.
That is because the Fermi level in bcc Fe sits at the bottom of the pseudo gap in the minority DOS (where $N_{\downarrow}$ is small) and at the shoulder of the majority-spin band dominated by $t_{2\text{g}\uparrow}$ states (where $N_{\uparrow}$ is relatively high)~\cite{Zhang:2010,Drittler:1989}.
In the Al, Co, and Ni containing binaries, alloying increases mainly $E_{\text{band}\uparrow}$ exceeding significantly the nearly concentration-independent contribution of $E_{\text{band}\downarrow}$.
With V, Cr, or Mn addition, alloying still increases $E_{\text{band}\uparrow}$ much more significantly than $E_{\text{band}\downarrow}$ but the main contribution to the increase of the band energy remains the minority-spin band, like in Fe.

The common denominator for the alloying-induced increase of $E_{\text{band}\uparrow}$ is basically due to a downshift of the Fe host states in the majority-spin band. The inset of Fig.~\ref{fig:DOS} illustrates the lowering of the edge of the $t_{2\text{g}\uparrow}$ shoulder with respect to $e_{\text{F}}$. While the edge of the $t_{2\text{g}\uparrow}$ shoulder is located just above $e_{\text{F}}$ in Fe, it is pushed below $e_{\text{F}}$ for all binaries, indicating the opening of a small Stoner pseudo-gap.
The associated microscopic mechanism in the case of Al, V, Cr, and Mn doping originates from the hybridization of the solute states with the unoccupied $s$ and $p$ bands of Fe~\cite{Drittler:1989}. In the case of Co and Ni, the solutes fill up mainly the majority-spin band of Fe. Both mechanisms lower $N_\uparrow(e_{\text{F}})$, decrease the magnetic susceptibility and make the ferromagnetism stronger relative to that of pure Fe. Stronger ferromagnetism, in turn, corresponds to more robust magnetic moments, to smaller excess magnetic pressure upon lattice deformation and to smaller difference between $\sigma^{\Omega/\mu}_{\text{m}}$ (Fig.~\ref{fig:SED}) and $\sigma_\text{m}$ (Fig.~\ref{fig:ITS}).

In summary, we have demonstrated that alloying Fe with frequently utilized solutes with concentrations up to 10\,\% is an effective mean to alter the intrinsic upper bound of the mechanical strength in tension along $\langle 001 \rangle$.
Vanadium turns out to be one of the most efficient alloying agents producing an enhancement of the ideal tensile strength by 2.3\,\% per atomic percent of V.
All binary systems considered here fail by cleavage under $\langle 001 \rangle$ loading.
The predicted ITSs form a consistent starting point for establishing useful limits on the attainable combination of strength and toughness of Fe-based alloys~\cite{Guo:2001,*Morris:2001b}.

We have shown that both the anomalously low ITS of pure Fe, in comparison with other bcc elements exhibiting the same failure mode, and the unexpected alloying effects are ascribed to the weak ferromagnetism of Fe. We have found that the present solutes alter the magnetic response of the Fe host during tension from the weak towards a stronger ferromagnetic behavior. The underlying driving force is shown to originate from the alloying induced-effects on the peculiar electronic structure of Fe. The fact that a small amount of Al or early $3d$ metal enhances the stability of ferromagnetism in Fe is in contrast to the commonly accepted scenario based on the phenomenological Slater-Pauling curve~\cite{Mohn:2006,*Acet:2010}, and calls for revision of the existing picture of magnetism in dilute Fe-alloys.

\emph{Acknowledgements} The Swedish Research Council, the Swedish Steel Producers' Association, the European Research Council, the China Scholarship Council, the Hungarian Scientific Research Fund (research projects OTKA 84078 and 109570), and the National Magnetic Confinement Fusion Program of China (2011GB108007) are acknowledged for financial support. The computations were performed using resources provided by the Swedish National Infrastructure for Computing (SNIC) at the National Supercomputer Centre in Link\"oping.


\begin{thebibliography}{46}%
\makeatletter
\providecommand \@ifxundefined [1]{%
 \@ifx{#1\undefined}
}%
\providecommand \@ifnum [1]{%
 \ifnum #1\expandafter \@firstoftwo
 \else \expandafter \@secondoftwo
 \fi
}%
\providecommand \@ifx [1]{%
 \ifx #1\expandafter \@firstoftwo
 \else \expandafter \@secondoftwo
 \fi
}%
\providecommand \natexlab [1]{#1}%
\providecommand \enquote  [1]{``#1''}%
\providecommand \bibnamefont  [1]{#1}%
\providecommand \bibfnamefont [1]{#1}%
\providecommand \citenamefont [1]{#1}%
\providecommand \href@noop [0]{\@secondoftwo}%
\providecommand \href [0]{\begingroup \@sanitize@url \@href}%
\providecommand \@href[1]{\@@startlink{#1}\@@href}%
\providecommand \@@href[1]{\endgroup#1\@@endlink}%
\providecommand \@sanitize@url [0]{\catcode `\\12\catcode `\$12\catcode
  `\&12\catcode `\#12\catcode `\^12\catcode `\_12\catcode `\%12\relax}%
\providecommand \@@startlink[1]{}%
\providecommand \@@endlink[0]{}%
\providecommand \url  [0]{\begingroup\@sanitize@url \@url }%
\providecommand \@url [1]{\endgroup\@href {#1}{\urlprefix }}%
\providecommand \urlprefix  [0]{URL }%
\providecommand \Eprint [0]{\href }%
\providecommand \doibase [0]{http://dx.doi.org/}%
\providecommand \selectlanguage [0]{\@gobble}%
\providecommand \bibinfo  [0]{\@secondoftwo}%
\providecommand \bibfield  [0]{\@secondoftwo}%
\providecommand \translation [1]{[#1]}%
\providecommand \BibitemOpen [0]{}%
\providecommand \bibitemStop [0]{}%
\providecommand \bibitemNoStop [0]{.\EOS\space}%
\providecommand \EOS [0]{\spacefactor3000\relax}%
\providecommand \BibitemShut  [1]{\csname bibitem#1\endcsname}%
\let\auto@bib@innerbib\@empty
\bibitem [{\citenamefont {Kelly}\ and\ \citenamefont
  {Macmillan}(1986)}]{Kelly:1986}%
  \BibitemOpen
  \bibfield  {author} {\bibinfo {author} {\bibfnamefont {A.}~\bibnamefont
  {Kelly}}\ and\ \bibinfo {author} {\bibfnamefont {N.~H.}\ \bibnamefont
  {Macmillan}},\ }\href@noop {} {\emph {\bibinfo {title} {Strong Solids}}}\
  (\bibinfo  {publisher} {Clarendon},\ \bibinfo {address} {Oxford},\ \bibinfo
  {year} {1986})\BibitemShut {NoStop}%
\bibitem [{\citenamefont {Clatterbuck}\ \emph
  {et~al.}(2003{\natexlab{a}})\citenamefont {Clatterbuck}, \citenamefont
  {Chrzan},\ and\ \citenamefont {{Morris, Jr.}}}]{inherent:property}%
  \BibitemOpen
  \bibfield  {author} {\bibinfo {author} {\bibfnamefont {D.~M.}\ \bibnamefont
  {Clatterbuck}}, \bibinfo {author} {\bibfnamefont {D.~C.}\ \bibnamefont
  {Chrzan}}, \ and\ \bibinfo {author} {\bibfnamefont {J.~W.}\ \bibnamefont
  {{Morris, Jr.}}},\ }\href@noop {} {\bibfield  {journal} {\bibinfo  {journal}
  {Acta Mater.}\ }\textbf {\bibinfo {volume} {51}},\ \bibinfo {pages} {2271}
  (\bibinfo {year} {2003}{\natexlab{a}})}\BibitemShut {NoStop}%
\bibitem [{\citenamefont {Nagasako}\ \emph {et~al.}(2010)\citenamefont
  {Nagasako}, \citenamefont {Jahn\'{a}tek}, \citenamefont {Asahi},\ and\
  \citenamefont {Hafner}}]{study:8}%
  \BibitemOpen
  \bibfield  {author} {\bibinfo {author} {\bibfnamefont {N.}~\bibnamefont
  {Nagasako}}, \bibinfo {author} {\bibfnamefont {M.}~\bibnamefont
  {Jahn\'{a}tek}}, \bibinfo {author} {\bibfnamefont {R.}~\bibnamefont {Asahi}},
  \ and\ \bibinfo {author} {\bibfnamefont {J.}~\bibnamefont {Hafner}},\
  }\href@noop {} {\bibfield  {journal} {\bibinfo  {journal} {Phys. Rev. B}\
  }\textbf {\bibinfo {volume} {81}},\ \bibinfo {pages} {094108} (\bibinfo
  {year} {2010})}\BibitemShut {NoStop}%
\bibitem [{\citenamefont {Li}\ \emph {et~al.}(2007)\citenamefont {Li},
  \citenamefont {{Morris, Jr.}}, \citenamefont {Nagasako}, \citenamefont
  {Kuramoto},\ and\ \citenamefont {Chrzan}}]{Li:2007}%
  \BibitemOpen
  \bibfield  {author} {\bibinfo {author} {\bibfnamefont {T.}~\bibnamefont
  {Li}}, \bibinfo {author} {\bibfnamefont {J.~W.}\ \bibnamefont {{Morris,
  Jr.}}}, \bibinfo {author} {\bibfnamefont {N.}~\bibnamefont {Nagasako}},
  \bibinfo {author} {\bibfnamefont {S.}~\bibnamefont {Kuramoto}}, \ and\
  \bibinfo {author} {\bibfnamefont {D.~C.}\ \bibnamefont {Chrzan}},\
  }\href@noop {} {\bibfield  {journal} {\bibinfo  {journal} {Phys. Rev. Lett.}\
  }\textbf {\bibinfo {volume} {98}},\ \bibinfo {pages} {105503} (\bibinfo
  {year} {2007})}\BibitemShut {NoStop}%
\bibitem [{\citenamefont {Qi}\ and\ \citenamefont {Chrzan}(2014)}]{Qi:2014}%
  \BibitemOpen
  \bibfield  {author} {\bibinfo {author} {\bibfnamefont {L.}~\bibnamefont
  {Qi}}\ and\ \bibinfo {author} {\bibfnamefont {D.~C.}\ \bibnamefont
  {Chrzan}},\ }\href@noop {} {\bibfield  {journal} {\bibinfo  {journal} {Phys.
  Rev. Lett.}\ }\textbf {\bibinfo {volume} {112}},\ \bibinfo {pages} {115503}
  (\bibinfo {year} {2014})}\BibitemShut {NoStop}%
\bibitem [{\citenamefont {Thomson}(1986)}]{Thomson:1986}%
  \BibitemOpen
  \bibfield  {author} {\bibinfo {author} {\bibfnamefont {R.}~\bibnamefont
  {Thomson}},\ }in\ \href@noop {} {\emph {\bibinfo {booktitle} {{Solid State
  Physics}}}},\ Vol.~\bibinfo {volume} {39},\ \bibinfo {editor} {edited by\
  \bibinfo {editor} {\bibfnamefont {H.}~\bibnamefont {Ehrenreich}}\ and\
  \bibinfo {editor} {\bibfnamefont {D.}~\bibnamefont {Turnbull}}}\ (\bibinfo
  {publisher} {Academic Press},\ \bibinfo {address} {New York},\ \bibinfo
  {year} {1986})\ p.~\bibinfo {pages} {1}\BibitemShut {NoStop}%
\bibitem [{\citenamefont {Jokl}\ \emph {et~al.}(1980)\citenamefont {Jokl},
  \citenamefont {Vitek},\ and\ \citenamefont {McMahon}}]{Jokl:1980}%
  \BibitemOpen
  \bibfield  {author} {\bibinfo {author} {\bibfnamefont {M.~J.}\ \bibnamefont
  {Jokl}}, \bibinfo {author} {\bibfnamefont {V.}~\bibnamefont {Vitek}}, \ and\
  \bibinfo {author} {\bibfnamefont {C.~J.}\ \bibnamefont {McMahon}},\
  }\href@noop {} {\bibfield  {journal} {\bibinfo  {journal} {Acta Metall.}\
  }\textbf {\bibinfo {volume} {28}},\ \bibinfo {pages} {1479} (\bibinfo {year}
  {1980})}\BibitemShut {NoStop}%
\bibitem [{\citenamefont {Krenn}\ \emph {et~al.}(2002)\citenamefont {Krenn},
  \citenamefont {Roundy}, \citenamefont {Cohen}, \citenamefont {Chrzan},\ and\
  \citenamefont {{Morris, Jr.}}}]{Krenn:2002}%
  \BibitemOpen
  \bibfield  {author} {\bibinfo {author} {\bibfnamefont {C.~R.}\ \bibnamefont
  {Krenn}}, \bibinfo {author} {\bibfnamefont {D.}~\bibnamefont {Roundy}},
  \bibinfo {author} {\bibfnamefont {M.~L.}\ \bibnamefont {Cohen}}, \bibinfo
  {author} {\bibfnamefont {D.~C.}\ \bibnamefont {Chrzan}}, \ and\ \bibinfo
  {author} {\bibfnamefont {J.~W.}\ \bibnamefont {{Morris, Jr.}}},\ }\href@noop
  {} {\bibfield  {journal} {\bibinfo  {journal} {Phys. Rev. B}\ }\textbf
  {\bibinfo {volume} {65}},\ \bibinfo {pages} {134111} (\bibinfo {year}
  {2002})}\BibitemShut {NoStop}%
\bibitem [{\citenamefont {Lee}\ \emph {et~al.}(2008)\citenamefont {Lee},
  \citenamefont {Wei}, \citenamefont {Kysar},\ and\ \citenamefont
  {Hone}}]{Lee:2008}%
  \BibitemOpen
  \bibfield  {author} {\bibinfo {author} {\bibfnamefont {C.}~\bibnamefont
  {Lee}}, \bibinfo {author} {\bibfnamefont {X.}~\bibnamefont {Wei}}, \bibinfo
  {author} {\bibfnamefont {J.~W.}\ \bibnamefont {Kysar}}, \ and\ \bibinfo
  {author} {\bibfnamefont {J.}~\bibnamefont {Hone}},\ }\href@noop {} {\bibfield
   {journal} {\bibinfo  {journal} {Science}\ }\textbf {\bibinfo {volume}
  {321}},\ \bibinfo {pages} {385} (\bibinfo {year} {2008})}\BibitemShut
  {NoStop}%
\bibitem [{\citenamefont {Roundy}\ \emph {et~al.}(1999)\citenamefont {Roundy},
  \citenamefont {Krenn}, \citenamefont {Cohen},\ and\ \citenamefont {{Morris,
  Jr.}}}]{study:9}%
  \BibitemOpen
  \bibfield  {author} {\bibinfo {author} {\bibfnamefont {D.}~\bibnamefont
  {Roundy}}, \bibinfo {author} {\bibfnamefont {C.~R.}\ \bibnamefont {Krenn}},
  \bibinfo {author} {\bibfnamefont {M.~L.}\ \bibnamefont {Cohen}}, \ and\
  \bibinfo {author} {\bibfnamefont {J.~W.}\ \bibnamefont {{Morris, Jr.}}},\
  }\href@noop {} {\bibfield  {journal} {\bibinfo  {journal} {Phys. Rev. Lett.}\
  }\textbf {\bibinfo {volume} {82}},\ \bibinfo {pages} {2713} (\bibinfo {year}
  {1999})}\BibitemShut {NoStop}%
\bibitem [{\citenamefont {Clatterbuck}\ \emph
  {et~al.}(2003{\natexlab{b}})\citenamefont {Clatterbuck}, \citenamefont
  {Krenn}, \citenamefont {Cohen},\ and\ \citenamefont {{Morris,
  Jr.}}}]{Clatterbuck:2003}%
  \BibitemOpen
  \bibfield  {author} {\bibinfo {author} {\bibfnamefont {D.~M.}\ \bibnamefont
  {Clatterbuck}}, \bibinfo {author} {\bibfnamefont {C.~R.}\ \bibnamefont
  {Krenn}}, \bibinfo {author} {\bibfnamefont {M.~L.}\ \bibnamefont {Cohen}}, \
  and\ \bibinfo {author} {\bibfnamefont {J.~W.}\ \bibnamefont {{Morris,
  Jr.}}},\ }\href@noop {} {\bibfield  {journal} {\bibinfo  {journal} {Phys.
  Rev. Lett.}\ }\textbf {\bibinfo {volume} {91}},\ \bibinfo {pages} {135501}
  (\bibinfo {year} {2003}{\natexlab{b}})}\BibitemShut {NoStop}%
\bibitem [{\citenamefont {\v{C}ern\'{y}}\ \emph {et~al.}(2013)\citenamefont
  {\v{C}ern\'{y}}, \citenamefont {\v{S}est\'{a}k}, \citenamefont {Pokluda},\
  and\ \citenamefont {\v{S}ob}}]{Cerny:2013}%
  \BibitemOpen
  \bibfield  {author} {\bibinfo {author} {\bibfnamefont {M.}~\bibnamefont
  {\v{C}ern\'{y}}}, \bibinfo {author} {\bibfnamefont {P.}~\bibnamefont
  {\v{S}est\'{a}k}}, \bibinfo {author} {\bibfnamefont {J.}~\bibnamefont
  {Pokluda}}, \ and\ \bibinfo {author} {\bibfnamefont {M.}~\bibnamefont
  {\v{S}ob}},\ }\href@noop {} {\bibfield  {journal} {\bibinfo  {journal} {Phys.
  Rev. B}\ }\textbf {\bibinfo {volume} {87}},\ \bibinfo {pages} {014117}
  (\bibinfo {year} {2013})}\BibitemShut {NoStop}%
\bibitem [{\citenamefont {Chen}\ \emph {et~al.}(2007)\citenamefont {Chen},
  \citenamefont {Gong},\ and\ \citenamefont {Wei}}]{Chen:2007}%
  \BibitemOpen
  \bibfield  {author} {\bibinfo {author} {\bibfnamefont {S.}~\bibnamefont
  {Chen}}, \bibinfo {author} {\bibfnamefont {X.~G.}\ \bibnamefont {Gong}}, \
  and\ \bibinfo {author} {\bibfnamefont {S.~H.}\ \bibnamefont {Wei}},\
  }\href@noop {} {\bibfield  {journal} {\bibinfo  {journal} {Phys. Rev. Lett.}\
  }\textbf {\bibinfo {volume} {98}},\ \bibinfo {pages} {015502} (\bibinfo
  {year} {2007})}\BibitemShut {NoStop}%
\bibitem [{\citenamefont {Zhang}\ \emph {et~al.}(2004)\citenamefont {Zhang},
  \citenamefont {Sun},\ and\ \citenamefont {Chen}}]{Zhang:2004}%
  \BibitemOpen
  \bibfield  {author} {\bibinfo {author} {\bibfnamefont {Y.}~\bibnamefont
  {Zhang}}, \bibinfo {author} {\bibfnamefont {H.}~\bibnamefont {Sun}}, \ and\
  \bibinfo {author} {\bibfnamefont {C.}~\bibnamefont {Chen}},\ }\href@noop {}
  {\bibfield  {journal} {\bibinfo  {journal} {Phys. Rev. Lett.}\ }\textbf
  {\bibinfo {volume} {93}},\ \bibinfo {pages} {195504} (\bibinfo {year}
  {2004})}\BibitemShut {NoStop}%
\bibitem [{\citenamefont {Blase}\ \emph
  {et~al.}(2004{\natexlab{a}})\citenamefont {Blase}, \citenamefont {Gillet},
  \citenamefont {San~Miguel},\ and\ \citenamefont {M\'elinon}}]{Blase:2004}%
  \BibitemOpen
  \bibfield  {author} {\bibinfo {author} {\bibfnamefont {X.}~\bibnamefont
  {Blase}}, \bibinfo {author} {\bibfnamefont {P.}~\bibnamefont {Gillet}},
  \bibinfo {author} {\bibfnamefont {A.}~\bibnamefont {San~Miguel}}, \ and\
  \bibinfo {author} {\bibfnamefont {P.}~\bibnamefont {M\'elinon}},\ }\href@noop
  {} {\bibfield  {journal} {\bibinfo  {journal} {Phys. Rev. Lett.}\ }\textbf
  {\bibinfo {volume} {92}},\ \bibinfo {pages} {215505} (\bibinfo {year}
  {2004}{\natexlab{a}})}\BibitemShut {NoStop}%
\bibitem [{\citenamefont {Blase}\ \emph
  {et~al.}(2004{\natexlab{b}})\citenamefont {Blase}, \citenamefont {Gillet},
  \citenamefont {San~Miguel},\ and\ \citenamefont {M\'elinon}}]{Blase:2}%
  \BibitemOpen
  \bibfield  {author} {\bibinfo {author} {\bibfnamefont {X.}~\bibnamefont
  {Blase}}, \bibinfo {author} {\bibfnamefont {P.}~\bibnamefont {Gillet}},
  \bibinfo {author} {\bibfnamefont {A.}~\bibnamefont {San~Miguel}}, \ and\
  \bibinfo {author} {\bibfnamefont {P.}~\bibnamefont {M\'elinon}},\ }\href
  {\doibase 10.1103/PhysRevLett.93.239901} {\bibfield  {journal} {\bibinfo
  {journal} {Phys. Rev. Lett.}\ }\textbf {\bibinfo {volume} {93}},\ \bibinfo
  {pages} {239901} (\bibinfo {year} {2004}{\natexlab{b}})}\BibitemShut
  {NoStop}%
\bibitem [{\citenamefont {Jiang}\ and\ \citenamefont
  {Srinivasan}(2013)}]{jiang:2013}%
  \BibitemOpen
  \bibfield  {author} {\bibinfo {author} {\bibfnamefont {C.}~\bibnamefont
  {Jiang}}\ and\ \bibinfo {author} {\bibfnamefont {S.~G.}\ \bibnamefont
  {Srinivasan}},\ }\href@noop {} {\bibfield  {journal} {\bibinfo  {journal}
  {Nature}\ }\textbf {\bibinfo {volume} {496}},\ \bibinfo {pages} {339}
  (\bibinfo {year} {2013})}\BibitemShut {NoStop}%
\bibitem [{\citenamefont {Wang}\ and\ \citenamefont {Wang}(2009)}]{study:1}%
  \BibitemOpen
  \bibfield  {author} {\bibinfo {author} {\bibfnamefont {Y.~J.}\ \bibnamefont
  {Wang}}\ and\ \bibinfo {author} {\bibfnamefont {C.~Y.}\ \bibnamefont
  {Wang}},\ }\href@noop {} {\bibfield  {journal} {\bibinfo  {journal} {Appl.
  Phys. Lett.}\ }\textbf {\bibinfo {volume} {94}},\ \bibinfo {pages} {261909}
  (\bibinfo {year} {2009})}\BibitemShut {NoStop}%
\bibitem [{\citenamefont {Guo}(2001)}]{Guo:2001}%
  \BibitemOpen
  \bibfield  {author} {\bibinfo {author} {\bibfnamefont {Z.}~\bibnamefont
  {Guo}},\ }\emph {\bibinfo {title} {The Limit of Strength and Toughness of
  Steel}},\ \href@noop {} {Ph.D. thesis},\ \bibinfo  {school} {University of
  California, Berkely} (\bibinfo {year} {2001})\BibitemShut {NoStop}%
\bibitem [{\citenamefont {{Morris, Jr.}}\ \emph {et~al.}(2001)\citenamefont
  {{Morris, Jr.}}, \citenamefont {Guo}, \citenamefont {Krenn},\ and\
  \citenamefont {Kim}}]{Morris:2001b}%
  \BibitemOpen
  \bibfield  {author} {\bibinfo {author} {\bibfnamefont {J.~W.}\ \bibnamefont
  {{Morris, Jr.}}}, \bibinfo {author} {\bibfnamefont {Z.}~\bibnamefont {Guo}},
  \bibinfo {author} {\bibfnamefont {C.~R.}\ \bibnamefont {Krenn}}, \ and\
  \bibinfo {author} {\bibfnamefont {Y.~H.}\ \bibnamefont {Kim}},\ }\href@noop
  {} {\bibfield  {journal} {\bibinfo  {journal} {ISIJ international}\ }\textbf
  {\bibinfo {volume} {41}},\ \bibinfo {pages} {599} (\bibinfo {year}
  {2001})}\BibitemShut {NoStop}%
\bibitem [{\citenamefont {\v{C}ern\'{y}}\ and\ \citenamefont
  {Pokluda}(2007)}]{Cerny:2007}%
  \BibitemOpen
  \bibfield  {author} {\bibinfo {author} {\bibfnamefont {M.}~\bibnamefont
  {\v{C}ern\'{y}}}\ and\ \bibinfo {author} {\bibfnamefont {J.}~\bibnamefont
  {Pokluda}},\ }\href@noop {} {\bibfield  {journal} {\bibinfo  {journal} {Phys.
  Rev. B}\ }\textbf {\bibinfo {volume} {76}},\ \bibinfo {pages} {024115}
  (\bibinfo {year} {2007})}\BibitemShut {NoStop}%
\bibitem [{\citenamefont {\v{S}ob}\ \emph {et~al.}(2004)\citenamefont
  {\v{S}ob}, \citenamefont {Fri\'ak}, \citenamefont {Legut}, \citenamefont
  {Fiala},\ and\ \citenamefont {Vitek}}]{Sob:2004}%
  \BibitemOpen
  \bibfield  {author} {\bibinfo {author} {\bibfnamefont {M.}~\bibnamefont
  {\v{S}ob}}, \bibinfo {author} {\bibfnamefont {M.}~\bibnamefont {Fri\'ak}},
  \bibinfo {author} {\bibfnamefont {D.}~\bibnamefont {Legut}}, \bibinfo
  {author} {\bibfnamefont {J.}~\bibnamefont {Fiala}}, \ and\ \bibinfo {author}
  {\bibfnamefont {V.}~\bibnamefont {Vitek}},\ }\href@noop {} {\bibfield
  {journal} {\bibinfo  {journal} {Mat. Sci. Eng. A}\ }\textbf {\bibinfo
  {volume} {387-389}},\ \bibinfo {pages} {148} (\bibinfo {year}
  {2004})}\BibitemShut {NoStop}%
\bibitem [{\citenamefont {\v{C}ern\'{y}}\ and\ \citenamefont
  {Pokluda}(2010)}]{Cerny:2010}%
  \BibitemOpen
  \bibfield  {author} {\bibinfo {author} {\bibfnamefont {M.}~\bibnamefont
  {\v{C}ern\'{y}}}\ and\ \bibinfo {author} {\bibfnamefont {J.}~\bibnamefont
  {Pokluda}},\ }\href@noop {} {\bibfield  {journal} {\bibinfo  {journal} {Phys.
  Rev. B}\ }\textbf {\bibinfo {volume} {82}},\ \bibinfo {pages} {174106}
  (\bibinfo {year} {2010})}\BibitemShut {NoStop}%
\bibitem [{\citenamefont {Luo}\ \emph {et~al.}(2002)\citenamefont {Luo},
  \citenamefont {Roundy}, \citenamefont {Cohen},\ and\ \citenamefont {{Morris,
  Jr.}}}]{Mo:1}%
  \BibitemOpen
  \bibfield  {author} {\bibinfo {author} {\bibfnamefont {W.}~\bibnamefont
  {Luo}}, \bibinfo {author} {\bibfnamefont {D.}~\bibnamefont {Roundy}},
  \bibinfo {author} {\bibfnamefont {M.~L.}\ \bibnamefont {Cohen}}, \ and\
  \bibinfo {author} {\bibfnamefont {J.~W.}\ \bibnamefont {{Morris, Jr.}}},\
  }\href@noop {} {\bibfield  {journal} {\bibinfo  {journal} {Phys. Rev. B}\
  }\textbf {\bibinfo {volume} {66}},\ \bibinfo {pages} {094110} (\bibinfo
  {year} {2002})}\BibitemShut {NoStop}%
\bibitem [{\citenamefont {Clatterbuck}\ \emph {et~al.}(2002)\citenamefont
  {Clatterbuck}, \citenamefont {Chrzan},\ and\ \citenamefont {{Morris,
  Jr.}}}]{idealandmagnetic:2002}%
  \BibitemOpen
  \bibfield  {author} {\bibinfo {author} {\bibfnamefont {D.~M.}\ \bibnamefont
  {Clatterbuck}}, \bibinfo {author} {\bibfnamefont {D.~C.}\ \bibnamefont
  {Chrzan}}, \ and\ \bibinfo {author} {\bibfnamefont {J.~W.}\ \bibnamefont
  {{Morris, Jr.}}},\ }\href@noop {} {\bibfield  {journal} {\bibinfo  {journal}
  {Phil. Mag. Lett.}\ }\textbf {\bibinfo {volume} {82}},\ \bibinfo {pages}
  {141} (\bibinfo {year} {2002})}\BibitemShut {NoStop}%
\bibitem [{\citenamefont {L.Tsetseris}(2005)}]{Tsetseris:2005}%
  \BibitemOpen
  \bibfield  {author} {\bibinfo {author} {\bibnamefont {L.Tsetseris}},\
  }\href@noop {} {\bibfield  {journal} {\bibinfo  {journal} {Phys. Rev. B}\
  }\textbf {\bibinfo {volume} {72}},\ \bibinfo {pages} {012411} (\bibinfo
  {year} {2005})}\BibitemShut {NoStop}%
\bibitem [{\citenamefont {Fri\'{a}k}\ \emph {et~al.}(2001)\citenamefont
  {Fri\'{a}k}, \citenamefont {\v{S}ob},\ and\ \citenamefont
  {Vitek}}]{Friak:2001}%
  \BibitemOpen
  \bibfield  {author} {\bibinfo {author} {\bibfnamefont {M.}~\bibnamefont
  {Fri\'{a}k}}, \bibinfo {author} {\bibfnamefont {M.}~\bibnamefont {\v{S}ob}},
  \ and\ \bibinfo {author} {\bibfnamefont {V.}~\bibnamefont {Vitek}},\
  }\href@noop {} {\bibfield  {journal} {\bibinfo  {journal} {Phys. Rev. B}\
  }\textbf {\bibinfo {volume} {63}},\ \bibinfo {pages} {052405} (\bibinfo
  {year} {2001})}\BibitemShut {NoStop}%
\bibitem [{\citenamefont {Liu}\ \emph {et~al.}(2009)\citenamefont {Liu},
  \citenamefont {Zhang}, \citenamefont {Hong},\ and\ \citenamefont {Lu}}]{Liu}%
  \BibitemOpen
  \bibfield  {author} {\bibinfo {author} {\bibfnamefont {Y.~L.}\ \bibnamefont
  {Liu}}, \bibinfo {author} {\bibfnamefont {Y.}~\bibnamefont {Zhang}}, \bibinfo
  {author} {\bibfnamefont {R.~J.}\ \bibnamefont {Hong}}, \ and\ \bibinfo
  {author} {\bibfnamefont {G.~H.}\ \bibnamefont {Lu}},\ }\href@noop {}
  {\bibfield  {journal} {\bibinfo  {journal} {Chin. Phys. B}\ }\textbf
  {\bibinfo {volume} {18}},\ \bibinfo {pages} {1923} (\bibinfo {year}
  {2009})}\BibitemShut {NoStop}%
\bibitem [{\citenamefont {Andersen}\ \emph {et~al.}(1994)\citenamefont
  {Andersen}, \citenamefont {Jepsen},\ and\ \citenamefont {Krier}}]{EMTO:1}%
  \BibitemOpen
  \bibfield  {author} {\bibinfo {author} {\bibfnamefont {O.~K.}\ \bibnamefont
  {Andersen}}, \bibinfo {author} {\bibfnamefont {O.}~\bibnamefont {Jepsen}}, \
  and\ \bibinfo {author} {\bibfnamefont {G.}~\bibnamefont {Krier}},\ }in\
  \href@noop {} {\emph {\bibinfo {booktitle} {{Lectures on Methods of
  Electronic Structure Calculations}}}},\ \bibinfo {editor} {edited by\
  \bibinfo {editor} {\bibfnamefont {V.}~\bibnamefont {Kumar}}, \bibinfo
  {editor} {\bibfnamefont {O.~K.}\ \bibnamefont {Andersen}}, \ and\ \bibinfo
  {editor} {\bibfnamefont {A.}~\bibnamefont {Mookerjee}}}\ (\bibinfo
  {publisher} {World Scientific},\ \bibinfo {address} {Singapore},\ \bibinfo
  {year} {1994})\ p.~\bibinfo {pages} {63}\BibitemShut {NoStop}%
\bibitem [{\citenamefont {Vitos}(2001)}]{EMTO:2}%
  \BibitemOpen
  \bibfield  {author} {\bibinfo {author} {\bibfnamefont {L.}~\bibnamefont
  {Vitos}},\ }\href@noop {} {\bibfield  {journal} {\bibinfo  {journal} {Phys.
  Rev. B}\ }\textbf {\bibinfo {volume} {64}},\ \bibinfo {pages} {014107}
  (\bibinfo {year} {2001})}\BibitemShut {NoStop}%
\bibitem [{\citenamefont {Vitos}\ \emph {et~al.}(2000)\citenamefont {Vitos},
  \citenamefont {Skriver}, \citenamefont {Johansson},\ and\ \citenamefont
  {Koll\'{a}r}}]{EMTO:3}%
  \BibitemOpen
  \bibfield  {author} {\bibinfo {author} {\bibfnamefont {L.}~\bibnamefont
  {Vitos}}, \bibinfo {author} {\bibfnamefont {H.~L.}\ \bibnamefont {Skriver}},
  \bibinfo {author} {\bibfnamefont {B.}~\bibnamefont {Johansson}}, \ and\
  \bibinfo {author} {\bibfnamefont {J.}~\bibnamefont {Koll\'{a}r}},\
  }\href@noop {} {\bibfield  {journal} {\bibinfo  {journal} {Comput. Mater.
  Sci.}\ }\textbf {\bibinfo {volume} {18}},\ \bibinfo {pages} {24} (\bibinfo
  {year} {2000})}\BibitemShut {NoStop}%
\bibitem [{\citenamefont {Perdew}\ \emph {et~al.}(1996)\citenamefont {Perdew},
  \citenamefont {Burke},\ and\ \citenamefont {Ernzerhof}}]{PBE}%
  \BibitemOpen
  \bibfield  {author} {\bibinfo {author} {\bibfnamefont {J.~P.}\ \bibnamefont
  {Perdew}}, \bibinfo {author} {\bibfnamefont {K.}~\bibnamefont {Burke}}, \
  and\ \bibinfo {author} {\bibfnamefont {M.}~\bibnamefont {Ernzerhof}},\
  }\href@noop {} {\bibfield  {journal} {\bibinfo  {journal} {Phys. Rev. Lett.}\
  }\textbf {\bibinfo {volume} {77}},\ \bibinfo {pages} {3865} (\bibinfo {year}
  {1996})}\BibitemShut {NoStop}%
\bibitem [{\citenamefont {Perdew}\ \emph {et~al.}(1997)\citenamefont {Perdew},
  \citenamefont {Burke},\ and\ \citenamefont {Ernzerhof}}]{Perdew:1996E}%
  \BibitemOpen
  \bibfield  {author} {\bibinfo {author} {\bibfnamefont {J.~P.}\ \bibnamefont
  {Perdew}}, \bibinfo {author} {\bibfnamefont {K.}~\bibnamefont {Burke}}, \
  and\ \bibinfo {author} {\bibfnamefont {M.}~\bibnamefont {Ernzerhof}},\
  }\href@noop {} {\bibfield  {journal} {\bibinfo  {journal} {Phys. Rev. Lett.}\
  }\textbf {\bibinfo {volume} {78}},\ \bibinfo {pages} {1396} (\bibinfo {year}
  {1997})}\BibitemShut {NoStop}%
\bibitem [{\citenamefont {Gy\H{o}rffy}(1972)}]{cpa:1}%
  \BibitemOpen
  \bibfield  {author} {\bibinfo {author} {\bibfnamefont {B.~L.}\ \bibnamefont
  {Gy\H{o}rffy}},\ }\href@noop {} {\bibfield  {journal} {\bibinfo  {journal}
  {Phys. Rev. B}\ }\textbf {\bibinfo {volume} {5}},\ \bibinfo {pages} {2382}
  (\bibinfo {year} {1972})}\BibitemShut {NoStop}%
\bibitem [{\citenamefont {Vitos}\ \emph {et~al.}(2001)\citenamefont {Vitos},
  \citenamefont {Abrikosov},\ and\ \citenamefont {Johansson}}]{cpa:3}%
  \BibitemOpen
  \bibfield  {author} {\bibinfo {author} {\bibfnamefont {L.}~\bibnamefont
  {Vitos}}, \bibinfo {author} {\bibfnamefont {I.~A.}\ \bibnamefont
  {Abrikosov}}, \ and\ \bibinfo {author} {\bibfnamefont {B.}~\bibnamefont
  {Johansson}},\ }\href@noop {} {\bibfield  {journal} {\bibinfo  {journal}
  {Phys. Rev. Lett.}\ }\textbf {\bibinfo {volume} {87}},\ \bibinfo {pages}
  {156401} (\bibinfo {year} {2001})}\BibitemShut {NoStop}%
\bibitem [{\citenamefont {Li}\ \emph {et~al.}(2013)\citenamefont {Li},
  \citenamefont {{Sch\"onecker}}, \citenamefont {Zhao}, \citenamefont
  {Johansson},\ and\ \citenamefont {Vitos}}]{Xiaoqing:2013}%
  \BibitemOpen
  \bibfield  {author} {\bibinfo {author} {\bibfnamefont {X.}~\bibnamefont
  {Li}}, \bibinfo {author} {\bibfnamefont {S.}~\bibnamefont {{Sch\"onecker}}},
  \bibinfo {author} {\bibfnamefont {J.}~\bibnamefont {Zhao}}, \bibinfo {author}
  {\bibfnamefont {B.}~\bibnamefont {Johansson}}, \ and\ \bibinfo {author}
  {\bibfnamefont {L.}~\bibnamefont {Vitos}},\ }\href@noop {} {\bibfield
  {journal} {\bibinfo  {journal} {Phys. Rev. B}\ }\textbf {\bibinfo {volume}
  {87}},\ \bibinfo {pages} {214203} (\bibinfo {year} {2013})}\BibitemShut
  {NoStop}%
\bibitem [{\citenamefont {Wills}\ \emph {et~al.}(1992)\citenamefont {Wills},
  \citenamefont {Eriksson}, \citenamefont {S\"oderlind},\ and\ \citenamefont
  {Boring}}]{Wills:1992}%
  \BibitemOpen
  \bibfield  {author} {\bibinfo {author} {\bibfnamefont {J.~M.}\ \bibnamefont
  {Wills}}, \bibinfo {author} {\bibfnamefont {O.}~\bibnamefont {Eriksson}},
  \bibinfo {author} {\bibfnamefont {P.}~\bibnamefont {S\"oderlind}}, \ and\
  \bibinfo {author} {\bibfnamefont {A.~M.}\ \bibnamefont {Boring}},\
  }\href@noop {} {\bibfield  {journal} {\bibinfo  {journal} {Phys.~Rev.~Lett.}\
  }\textbf {\bibinfo {volume} {68}},\ \bibinfo {pages} {2802} (\bibinfo {year}
  {1992})}\BibitemShut {NoStop}%
\bibitem [{\citenamefont {Souvatzis}\ \emph {et~al.}(2004)\citenamefont
  {Souvatzis}, \citenamefont {Katsnelson}, \citenamefont {Simak}, \citenamefont
  {Ahuja}, \citenamefont {Eriksson},\ and\ \citenamefont
  {Mohn}}]{Souvatzis:2004}%
  \BibitemOpen
  \bibfield  {author} {\bibinfo {author} {\bibfnamefont {P.}~\bibnamefont
  {Souvatzis}}, \bibinfo {author} {\bibfnamefont {M.~I.}\ \bibnamefont
  {Katsnelson}}, \bibinfo {author} {\bibfnamefont {S.}~\bibnamefont {Simak}},
  \bibinfo {author} {\bibfnamefont {R.}~\bibnamefont {Ahuja}}, \bibinfo
  {author} {\bibfnamefont {O.}~\bibnamefont {Eriksson}}, \ and\ \bibinfo
  {author} {\bibfnamefont {P.}~\bibnamefont {Mohn}},\ }\href@noop {} {\bibfield
   {journal} {\bibinfo  {journal} {Phys.~Rev.~B}\ }\textbf {\bibinfo {volume}
  {70}},\ \bibinfo {pages} {012201} (\bibinfo {year} {2004})}\BibitemShut
  {NoStop}%
\bibitem [{\citenamefont {Andersen}\ \emph {et~al.}(1977)\citenamefont
  {Andersen}, \citenamefont {Madsen}, \citenamefont {Poulsen}, \citenamefont
  {Jepsen},\ and\ \citenamefont {Koll\'ar}}]{Andersen:1977}%
  \BibitemOpen
  \bibfield  {author} {\bibinfo {author} {\bibfnamefont {O.~K.}\ \bibnamefont
  {Andersen}}, \bibinfo {author} {\bibfnamefont {J.}~\bibnamefont {Madsen}},
  \bibinfo {author} {\bibfnamefont {U.~K.}\ \bibnamefont {Poulsen}}, \bibinfo
  {author} {\bibfnamefont {O.}~\bibnamefont {Jepsen}}, \ and\ \bibinfo {author}
  {\bibfnamefont {J.}~\bibnamefont {Koll\'ar}},\ }\href@noop {} {\bibfield
  {journal} {\bibinfo  {journal} {Physica}\ }\textbf {\bibinfo {volume}
  {86-88B}},\ \bibinfo {pages} {249} (\bibinfo {year} {1977})}\BibitemShut
  {NoStop}%
\bibitem [{\citenamefont {Punkkinen}\ \emph {et~al.}(2011)\citenamefont
  {Punkkinen}, \citenamefont {Kwon}, \citenamefont {Koll\'{a}r}, \citenamefont
  {Johansson},\ and\ \citenamefont {Vitos}}]{Punkkinen:2011}%
  \BibitemOpen
  \bibfield  {author} {\bibinfo {author} {\bibfnamefont {M.~P.~J.}\
  \bibnamefont {Punkkinen}}, \bibinfo {author} {\bibfnamefont {S.~K.}\
  \bibnamefont {Kwon}}, \bibinfo {author} {\bibfnamefont {J.}~\bibnamefont
  {Koll\'{a}r}}, \bibinfo {author} {\bibfnamefont {B.}~\bibnamefont
  {Johansson}}, \ and\ \bibinfo {author} {\bibfnamefont {L.}~\bibnamefont
  {Vitos}},\ }\href@noop {} {\bibfield  {journal} {\bibinfo  {journal} {Phys.
  Rev. Lett.}\ }\textbf {\bibinfo {volume} {106}},\ \bibinfo {pages} {057202}
  (\bibinfo {year} {2011})}\BibitemShut {NoStop}%
\bibitem [{\citenamefont {Springford}(1980)}]{Springford:1980}%
  \BibitemOpen
  \bibinfo {editor} {\bibfnamefont {M.}~\bibnamefont {Springford}},\ ed.,\
  \enquote {\bibinfo {title} {{E}lectrons at the {F}ermi surface},}\ \
  (\bibinfo  {publisher} {Cambridge University Press},\ \bibinfo {address}
  {Cambridge},\ \bibinfo {year} {1980})\ Chap.~\bibinfo {chapter}
  {5}\BibitemShut {NoStop}%
\bibitem [{\citenamefont {Skriver}(1985)}]{Skriver:1985}%
  \BibitemOpen
  \bibfield  {author} {\bibinfo {author} {\bibfnamefont {H.~L.}\ \bibnamefont
  {Skriver}},\ }\href@noop {} {\bibfield  {journal} {\bibinfo  {journal}
  {Phys.~Rev.~B}\ }\textbf {\bibinfo {volume} {31}},\ \bibinfo {pages} {1909}
  (\bibinfo {year} {1985})}\BibitemShut {NoStop}%
\bibitem [{\citenamefont {Zhang}\ \emph {et~al.}(2010)\citenamefont {Zhang},
  \citenamefont {Punkkinen}, \citenamefont {Johansson}, \citenamefont
  {Hertzman},\ and\ \citenamefont {Vitos}}]{Zhang:2010}%
  \BibitemOpen
  \bibfield  {author} {\bibinfo {author} {\bibfnamefont {H.}~\bibnamefont
  {Zhang}}, \bibinfo {author} {\bibfnamefont {M.~P.~J.}\ \bibnamefont
  {Punkkinen}}, \bibinfo {author} {\bibfnamefont {B.}~\bibnamefont
  {Johansson}}, \bibinfo {author} {\bibfnamefont {S.}~\bibnamefont {Hertzman}},
  \ and\ \bibinfo {author} {\bibfnamefont {L.}~\bibnamefont {Vitos}},\
  }\href@noop {} {\bibfield  {journal} {\bibinfo  {journal} {Phys. Rev. B}\
  }\textbf {\bibinfo {volume} {81}},\ \bibinfo {pages} {184105} (\bibinfo
  {year} {2010})}\BibitemShut {NoStop}%
\bibitem [{\citenamefont {Drittler}\ \emph {et~al.}(1989)\citenamefont
  {Drittler}, \citenamefont {Stefanou}, \citenamefont {Bl\"ugel}, \citenamefont
  {Zeller},\ and\ \citenamefont {Dederichs}}]{Drittler:1989}%
  \BibitemOpen
  \bibfield  {author} {\bibinfo {author} {\bibfnamefont {B.}~\bibnamefont
  {Drittler}}, \bibinfo {author} {\bibfnamefont {N.}~\bibnamefont {Stefanou}},
  \bibinfo {author} {\bibfnamefont {S.}~\bibnamefont {Bl\"ugel}}, \bibinfo
  {author} {\bibfnamefont {R.}~\bibnamefont {Zeller}}, \ and\ \bibinfo {author}
  {\bibfnamefont {P.~H.}\ \bibnamefont {Dederichs}},\ }\href@noop {} {\bibfield
   {journal} {\bibinfo  {journal} {Phys. Rev. B}\ }\textbf {\bibinfo {volume}
  {40}},\ \bibinfo {pages} {8203} (\bibinfo {year} {1989})}\BibitemShut
  {NoStop}%
\bibitem [{\citenamefont {Mohn}(2006)}]{Mohn:2006}%
  \BibitemOpen
  \bibfield  {author} {\bibinfo {author} {\bibfnamefont {P.}~\bibnamefont
  {Mohn}},\ }\href@noop {} {\emph {\bibinfo {title} {{M}agnetism in the {S}olid
  {S}tate, {A}n {I}ntroduction}}},\ \bibinfo {edition} {2nd}\ ed.\ (\bibinfo
  {publisher} {Springer-Verlag},\ \bibinfo {address} {Berlin},\ \bibinfo {year}
  {2006})\BibitemShut {NoStop}%
\bibitem [{\citenamefont {Pepperhoff}\ and\ \citenamefont
  {Acet}(2010)}]{Acet:2010}%
  \BibitemOpen
  \bibfield  {author} {\bibinfo {author} {\bibfnamefont {W.}~\bibnamefont
  {Pepperhoff}}\ and\ \bibinfo {author} {\bibfnamefont {M.}~\bibnamefont
  {Acet}},\ }\href@noop {} {\emph {\bibinfo {title} {{C}onstitution and
  {M}agnetism of {I}ron and its {A}lloys}}}\ (\bibinfo  {publisher}
  {Springer},\ \bibinfo {address} {Berlin},\ \bibinfo {year}
  {2010})\BibitemShut {NoStop}%
\end{thebibliography}
\end{document}